\documentclass[reprint,aps,prb,superscriptaddress,amsmath,amssymb,showpacs,preprintnumbers,floatfix]{revtex4-1}

\usepackage{graphicx}
\usepackage{natbib,color,bm}
\usepackage{multirow}

\def\Qb{\ensuremath{\mathbf Q}}

\def\LRF{Li$R$F$_4$}
\def\Tc{\ensuremath{T_{\rm c}}}
\def\Hc{\ensuremath{H_{\rm c}}}
\def\TN{\ensuremath{T_{\rm N}}}

\begin{document}



\title{Neutron spectroscopic study of crystal-field excitations and the effect of the crystal field on dipolar magnetism in Li$R$F$_4$ ($R$ = Gd, Ho, Er, Tm, and Yb)}

\author{P.~Babkevich}
\email[]{peter.babkevich@epfl.ch}
\affiliation{Laboratory for Quantum Magnetism,  \'{E}cole Polytechnique F\'{e}d\'{e}rale de Lausanne (EPFL), CH-1015 Lausanne, Switzerland}
\author{A.~Finco}
\affiliation{Laboratory for Quantum Magnetism,  \'{E}cole Polytechnique F\'{e}d\'{e}rale de Lausanne (EPFL), CH-1015 Lausanne, Switzerland}
\affiliation{ICFP, D\'{e}partement de physique, \'{E}cole normale sup\'{e}rieure, 45 rue d'Ulm,
75005 Paris, France}
\author{M.~Jeong}
\affiliation{Laboratory for Quantum Magnetism,  \'{E}cole Polytechnique F\'{e}d\'{e}rale de Lausanne (EPFL), CH-1015 Lausanne, Switzerland}
\author{B.~Dalla Piazza}
\affiliation{Laboratory for Quantum Magnetism,  \'{E}cole Polytechnique F\'{e}d\'{e}rale de Lausanne (EPFL), CH-1015 Lausanne, Switzerland}
\author{I.~Kovacevic}
\affiliation{Laboratory for Quantum Magnetism,  \'{E}cole Polytechnique F\'{e}d\'{e}rale de Lausanne (EPFL), CH-1015 Lausanne, Switzerland}
\author{G.~Klughertz}
\affiliation{Laboratory for Quantum Magnetism,  \'{E}cole Polytechnique F\'{e}d\'{e}rale de Lausanne (EPFL), CH-1015 Lausanne, Switzerland}
\affiliation{Institut de Physique et Chimie des Mat\'{e}riaux de Strasbourg, CNRS and Universit\'{e} de Strasbourg, Bo\^{\i}te Postale 43, F-67034 Strasbourg, France}
\author{K.~W.~Kr\"{a}mer}
\affiliation{Department of Chemistry and Biochemistry, University of Bern, CH-3012 Bern, Switzerland}
\author{C.~Kraemer}
\affiliation{Laboratory for Quantum Magnetism,  \'{E}cole Polytechnique F\'{e}d\'{e}rale de Lausanne (EPFL), CH-1015 Lausanne, Switzerland}
\author{D.~T.~Adroja}
\affiliation{ISIS Facility, Rutherford Appleton Laboratory, Chilton, Didcot, Oxfordshire OX11 0QX, United Kingdom}
\author{E.~Goremychkin}
\affiliation{ISIS Facility, Rutherford Appleton Laboratory, Chilton, Didcot, Oxfordshire OX11 0QX, United Kingdom}
\author{T.~Unruh} 
\affiliation{Forschungsneutronenquelle Heinz Maier-Leibnitz (FRM II), Technische Universit\"{a}t M\"{u}nchen 85748 Garching, Germany}
\affiliation{Lehrstuhl f\"{u}r Kristallographie und Strukturphysik, Friedrich-Alexander-Universit\"{a}t Erlangen-N\"{u}rnberg, Physik Department - 91058 Erlangen, Germany}
\author{T.~Str\"{a}ssle}
\affiliation{Laboratory for Neutron Scattering, Paul Scherrer Institut, CH-5232 Villigen
PSI, Switzerland}
\author{A.~Di~Lieto}
\affiliation{Dipartimento di Fisica, Universita di Pisa, Largo B. Pontecorvo 3, I-56127 Pisa, Italy}
\affiliation{NEST Istituto di Nanoscienze-CNR, Piazza S. Silvestro 12, I-56127 Pisa, Italy}
\author{J. Jensen}
\affiliation{Niels Bohr Institute, Universitetsparken 5, 2100 Copenhagen, Denmark}
\author{H.~M.~R{\o}nnow}
\affiliation{Laboratory for Quantum Magnetism,  \'{E}cole Polytechnique F\'{e}d\'{e}rale de Lausanne (EPFL), CH-1015 Lausanne, Switzerland}

\date{\today}

\begin{abstract}
We present a systematic study of the crystal field interactions in the \LRF, $R =$ Gd, Ho, Er, Tm and Yb, family of rare-earth magnets. Using detailed inelastic neutron scattering measurements we have been able to quantify the transition energies and wavefunctions for each system. This allows us to quantitatively describe the high-temperature susceptibility measurements for the series of materials and make predictions based on a mean-field approach for the low-temperature thermal and quantum phase transitions. We show that coupling between crystal field and phonon states leads to lineshape broadening in LiTmF$_4$ and level splitting in LiYbF$_4$. Furthermore, using high-resolution neutron scattering from LiHoF$_4$, we find anomalous broadening of crystal-field excitations which we attribute to magnetoelastic coupling.
\end{abstract}



\maketitle

\graphicspath{{.}{Figures/}}

\section{Introduction}

Dipolar-coupled magnets are of great interest in the study of fundamental interactions and critical behavior. The family of \LRF\ insulating fluorides, where $R^{3+}$ is a rare-earth ion, displays a range of intriguing properties at low-temperatures. LiHoF$_4$ has attracted a great deal of interest owing to its Ising ferromagnetic order below 1.53\,K and a quantum phase transition in a transverse field of 5.0\,T.\cite{hansen-prb-1975, bitko-prl-1996, chakraborty-prb-2004, ronnow-science-2005, ronnow-prb-2007, tabei-prb-2008}
Diluted LiHo$_x$Y$_{1-x}$F$_4$ has been found to be an ideal material to investigate quantum phase transitions in a disordered system.\cite{schechter-prb-2008,tabei-prl-2006,andresen-prx-2014, nikseresht-unpub} While LiYF$_4$ is non-magnetic, it is frequently used in lasers as host material doped with trivalent rare-earth ions.\cite{chicklis-ieee-1972, kaminskii-book-1996} More recently, dimensional reduction was discovered in the XY antiferromagnet LiErF$_4$,\cite{kraemer-science-2012} and enhanced glassiness was reported for the mixed Ising-XY series LiHo$_x$Er$_{1-x}$F$_4$.\cite{piatek-prb-2014} Therefore, the family of \LRF\ allows us to probe a wide range of phenomena. Further progress to understand these systems, in particular close to quantum phase transitions, requires a detailed knowledge of the effect of the crystalline electric field (CEF) which typically determines the the ground state of the system.

\begin{table}
\centering
\begin{tabular*}{0.75\columnwidth}{@{\extracolsep{\fill} } l c c c c c c}
\hline
\hline
$R^{3+}$ & shell     &term               & $S$   & $L$ & $J$ & $g_J$ \\
\hline
Gd & $4f^7$     & $^8S_{7/2}$       & 7/2   & 0  & 7/2  & 2.00\\
Tb & $4f^8$     & $^7F_{6}$         & 3     & 3  & 6    & 1.50\\
Dy & $4f^9$     & $^6H_{15/2}$      & 5/2   & 5  & 15/2  & 1.33\\
Ho & $4f^{10}$  & $^5I_{8}$         & 2     & 6  & 8    & 1.25\\
Er & $4f^{11}$  & $^4I_{15/2}$      & 3/2   & 6  & 15/2  & 1.20\\
Tm & $4f^{12}$  & $^3H_{6}$         & 1     & 5  & 6    & 1.17\\
Yb & $4f^{13}$  & $^2F_{7/2}$       & 1/2    & 3  & 7/2  & 1.14\\
\hline
\hline
\end{tabular*}
\caption{Electronic configuration of the ground states of the heavier $4f$ ions. Hund's rules determine the values of $S$, $L$, and $J$. The corresponding Land\'{e} factors $g_J$ are shown in the last column.
\label{tab:SLJ}}
\end{table}

The electronic ground state configurations of $R^{3+}$ ions resulting from Hund's rules are shown in Table~\ref{tab:SLJ}. For more than half-filled shells the maximum $J=L+S$ state becomes the ground state. All states are further split by the CEF with a total splitting of 60\,meV for fluorides. The splitting may result in a maximum of $2J+1$ levels, depending on Kramers degeneracy and the point symmetry of the rare-earth site. Since the energy gap between the ground state multiplet and lowest excited electronic configuration is at least 3000\,K in \LRF\ compounds,\cite{jensen-book-1991} we only consider the ground state for the discussion of the magnetic properties.

The microscopic understanding of the \LRF\ systems is based upon a Hamiltonian of the following form,
\begin{equation}
\mathcal{H} = \mathcal{H}_{\rm CEF} + \mathcal{H}_{\rm D} + \mathcal{H}_{\rm Z} + \mathcal{H}_{\rm HF}+ \mathcal{H}_{\rm ex},
\end{equation}
where $\mathcal{H}_{\rm CEF}$ describes the crystalline electric field (CEF), $\mathcal{H}_{\rm D}$ the magnetic dipole-dipole interaction, $\mathcal{H}_{\rm Z}$ is the Zeeman term and the nearest-neighbor exchange interaction is given by $\mathcal{H}_{\rm ex}$. Additionally, some isotopes of $R$ carry a nuclear magnetic moment and a further hyperfine interaction $\mathcal{H}_{\rm HF}$ between the electronic and nuclear spin degrees of freedom may need to be taken into account.

Many studies have attempted to address the CEF environment of \LRF\ using Raman scattering, infrared spectroscopy and susceptibility measurements.\cite{jenssen-prb-1975, hansen-prb-1975, gifeisman-1978, christensen-prb-1979a, christensen-prb-1979b, beauvillain-jmmm-1980, gorller-book-1996, dong-pss-2003, abubakirov-jpcm-2008, feng-2009} However, there are a large range of values reported and only few studies have used inelastic neutron scattering as an experimental probe. In most cases neutron scattering allows one to tightly constrain the CEF parameters as neutrons can probe both the energies of the transitions between CEF levels and the transition matrix elements. In this paper we present high-resolution inelastic neutron scattering measurements of the CEF levels of \LRF\ for $R = $ Ho, Er, Yb and Tm. The extracted CEF parameters are refined using inelastic neutron and bulk magnetization measurements. Furthermore, the established Hamiltonians are used to predict the temperature-field phase diagrams of \LRF\ using a mean-field approximation.

\section{Structure and crystal field environment}
\label{sec:struct_cef}

\begin{figure}
\centering
\includegraphics[width=0.9\columnwidth,clip=]
{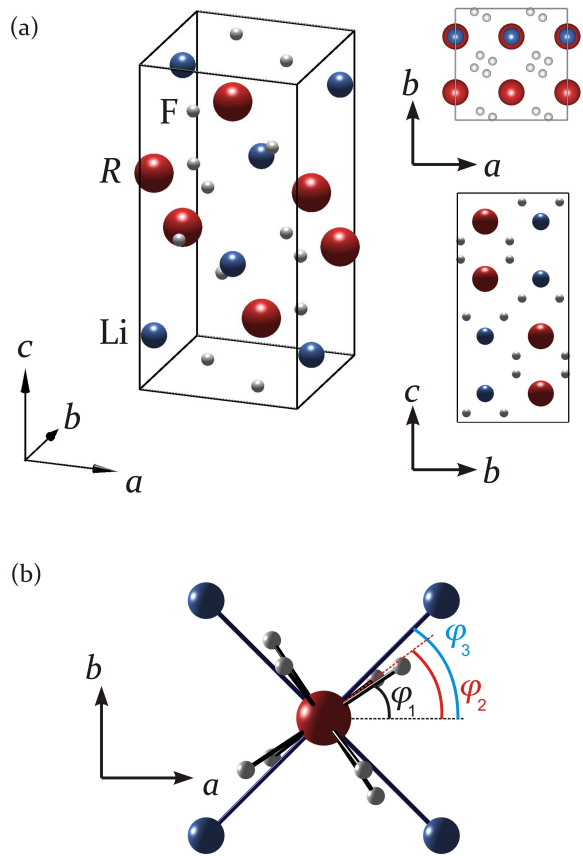}
\caption{(Color online) (a) Crystal structure of \LRF. Depending on $R$, the lattice parameters of the tetragonal structure, space group $I4_1/a$, $Z=4$ are approximately $a = b \simeq 5.2$\,\AA\ and $c \simeq 10.5$\,\AA. The atoms are situated at approximately: Li$^{+}$ (4a) $0,1/4,5/8$, $R^{3+}$ (4b) $0, 1/4, 1/8$ and F$^-$ (16f) $0.22, 0.58, 0.54$ sites for origin at $\bar{1}$.
(b) Coordination of $R$ ions by nearest eight F and four Li ions. The $R$-F and $R$-Li bond lengths shown are approximately 2.26\,\AA\ and 3.68\,\AA, respectively. In the $ab$ plane the ions subtend an angle of $\varphi_1 = 34^\circ$, $\varphi_2 = 37^\circ$ and $\varphi_3 = 45^\circ$ from the $a$ axis.
\label{fig:crystal_struct}}
\end{figure}

Compounds of the \LRF\ family crystalize in a tetragonal Scheelite lattice as shown in Fig.~\ref{fig:crystal_struct}. The Li and $R$ ions occupy $\bar{4}$-symmetry sites while the F ions are located at a general position such that none of them are on an inversion center. Minimal structural distortion is found when different $R$ ions are doped into the structure -- a small displacement of F ions and slight change in the lattice parameters.\cite{salaun-condmatt-1997a} The ratio of lattice parameters  $c/a$ has a linear dependence on the ionic radii of $R$.\cite{salaun-condmatt-1997a} It is also possible to dilute parent compounds with other rare-earths\cite{piatek-prb-2014} and non-magnetic ions such as Y and Lu.\cite{tabei-prl-2006, schechter-prb-2008}

The CEF determines the local magnetic properties of an ion such as the effective magnetic moment or the anisotropy. If there is no interaction with the neighboring ions, the ground state will be degenerate with $2J+1$ states. However, the CEF interaction lifts this degeneracy and in general it is strongly dependent on the symmetry of the crystallographic structure. The crystal-field Hamiltonian $\mathcal{H}_{\rm CEF}$ can then take a block-diagonal form where the multiplet is split into states that transform according to the one-dimensional irreducible representations of the $\bar{4}$ ($S_4$) point group. One can therefore consider subspaces that span the following states,
\begin{equation}
\Psi_m = \{ | -m \rangle , | - m + 4\rangle \ldots | m - 4\rangle , | m \rangle\},
\end{equation}
where $m=-J, \ldots ,J$. It can be shown that for electronic configurations that are parity-even and have integer $J$, the energy levels can be divided into three sets: two non-degenerate sets which transform according to $\Gamma_1$ and $\Gamma_2$ (when $m$ is even) representations and one doubly degenerate set transforming according to the $\Gamma_{3,4}$ (when $m$ is odd) representation of the $\bar{4}$ group and related by time-reversal symmetry. In this paper we employ the representation labeling convention defined by Koster.\cite{koster-book-1963} The matrix elements of the transverse angular momentum operators are zero for both the singlet and doublet states. In the case of a singlet state, time-reversal symmetry dictates that there cannot be a non-zero ${\mathbf J}_z$ matrix element. Therefore, the ions with integer $J$ cannot have planar anisotropy but are realizations of the Ising anisotropy along the $z$ direction ($c$ axis) with vanishing susceptibility in the transverse directions when the ground state is a doublet. A singlet ground state has no net magnetization at $T=0$.

In the case of half-integer $J$ the ions have odd parity and are governed by Kramers' theorem, which dictates that all states are doubly degenerate and divided into sets $\Gamma_{5,6}$ and $\Gamma_{7,8}$. They transform according to representations of the $\bar{4}$ double group. In this case it is possible to have planar spin anisotropy as the angular momentum operators ${\mathbf J}_x$ and ${\mathbf J}_y$ are allowed by symmetry to have non-zero matrix elements in the degenerate ground state subspace. For the rare-earth ions considered here, the CEF states decompose into irreducible representations as,\cite{herzig-book-1994, gorller-book-1996}
\begin{align}
&{\rm Ho}^{3+}&:& \quad 5\Gamma_1 + 4\Gamma_2 + 4\Gamma_{3,4}\nonumber\\
&{\rm Er}^{3+}&:& \quad 4\Gamma_{5,6} + 4\Gamma_{7,8}\nonumber\\
&{\rm Tm}^{3+}&:& \quad 3\Gamma_1 + 4\Gamma_2 + 3\Gamma_{3,4}\nonumber\\
&{\rm Yb}^{3+}&:& \quad 2\Gamma_{5,6} + 2\Gamma_{7,8}
\label{eq:ions}
\end{align}
The double subscripts refer to states which are degenerate and consist of a direct sum of two irreducible representations.

From the preceding discussion it is evident that the symmetry of the CEF environment restricts many of the properties of \LRF\ materials. The interaction of $R^{3+}$ with the surrounding Li$^+$ and F$^-$ ions is described by the effective crystal-field Hamiltonian which takes the general form of,
\begin{equation}
\mathcal{H}_{\rm CEF} = \sum_i \sum_{lm} B_l^m{\mathbf O}_l^m({\mathbf J}_i).
\label{eq:total_Ham_simple}
\end{equation}
The summation is taken over all the ions and we use the Stevens' operators ${\mathbf O}_l^m$ with energy coefficients $B_l^m$. We employ the convention where the ion-dependent Stevens factors are already included in our CEF parameters $B_l^m$. The total angular momentum operator is denoted by ${\mathbf J}$. The coefficients $B_l^m$ can be estimated from first-principles provided detailed knowledge of the surrounding charge distribution is known. However, in practice it is difficult to obtain reliable values and the coefficients are treated as free parameters which are determined experimentally.
To a first approximation we can define intrinsic $A^m_l$ CEF parameters which are independent of the particular lanthanide ion,\cite{hutchings-1964}
\begin{equation}
B^m_l = A^m_l \theta_l \langle r^l \rangle,
\end{equation}
where the Stevens factors $\theta_l$ depend on the form of the electronic charge cloud and $\langle r^l \rangle$ are the expectation values of the radial part of the $4f$ ion wavefunction.
The precise CEF Hamiltonian is uniquely defined by the local symmetry at the crystallographic site that the $R$ ion occupies. In \LRF\ the $R$ ion occupies the $4b$ site with $\bar{4}$ point symmetry which results in,
\begin{equation}
\mathcal{H}_{\rm CEF} = \sum_{l =2,4,6}  B_l^0{\mathbf O}_l^0 +
\sum_{l = 4,6}  B_l^4(c){\mathbf O}_l^4(c) + B_l^4(s){\mathbf O}_l^4(s).
\label{eq:crystal_field}
\end{equation}
We can apply a unitary transformation operator $\mathbf{U} = \exp(-{\rm i}\psi \mathbf{J}_z)$  on the Hamiltonian such that $\mathcal{H}^R_{\rm CEF} = \mathbf{U}^\dag\mathcal{H}_{\rm CEF} \mathbf{U}$.  Since CEF operators ${\mathbf O}_l^0$ commute with $\mathbf{U}$, the rotation has no effect on these terms. However, ${\mathbf O}_l^{\pm4}$ will transform on rotation and by choosing a suitable angle of rotation $\psi$ about the $z$ axis (along $c$ axis) we set $B^4_4(s) = 0$. We define a unique choice of CEF parameters by requiring that the $x$ axis is where $B_4^4(c)/\theta_4 < 0$ (or $A_4^4(c) < 0$) and $B^4_4(s) = 0$. Two possible equivalent coordinations of $R$ ion by F ions give different sign of $B_6^4(s)$. Consequently, most experiments leave the sign of $B_6^4(s)$ undetermined. Furthermore, coordination of the $R$ ions is close to the higher $\bar{4}2m$ ($D_{2d}$) symmetry and therefore we expect the parameter $B_6^4(s)$ to be small.

As will be shown later, the ${\mathbf O}_2^0$ term is the dominant CEF contribution in many of the \LRF\ systems. In order to minimize the energy of the system, planar anisotropy is preferred when $B_2^0 > 0$, while conversely when $B_2^0 < 0$ it is energetically more favorable to have the spins aligned along the $z$ axis. Hence, positive values of $B_l^0$, $l=2,4,6$ in the first sum of Eq.~\ref{eq:crystal_field} confine the magnetic moment to the $ab$ crystallographic plane. The coefficients in the second sum induce a small planar anisotropy.

\begin{figure}
\centering
\includegraphics[width=\columnwidth,clip=]
{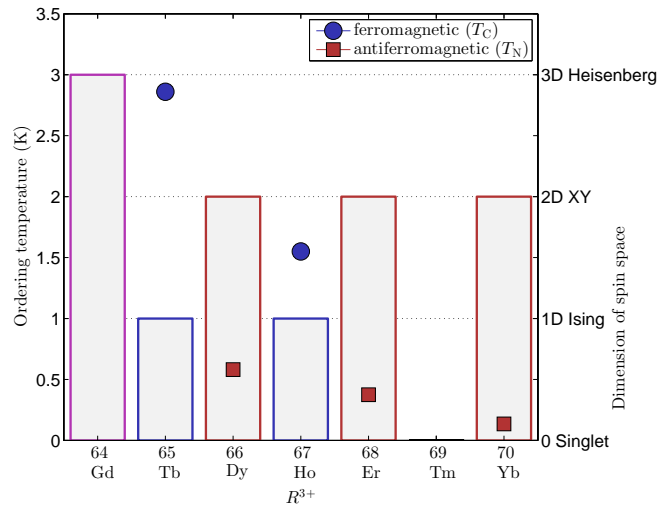}
\caption{(Color online) Magnetic properties of selected \LRF\ compounds. The bars denote the dimension of spin space while the (red squares) blue circles show the (anti)ferromagnetic ordering temperature. Data for transition temperatures are after Refs.~\onlinecite{aminov-book-1996, bitko-prl-1996, kraemer-science-2012}.
\label{fig:dimensionality}}
\end{figure}

The CEF together with the dipole-dipole interaction can lead to long-range magnetic order in the \LRF\ systems. Figure~\ref{fig:dimensionality} shows a summary of the magnetic properties of selected \LRF\ compounds with $R = {\rm Gd - Yb}$. Ferromagnetic order is found in LiTbF$_4$ and LiHoF$_4$ where in the case of Ho a non-Kramers doublet is realized while for Tb a quasi-doublet forms the ground state.\cite{aminov-book-1996} Systems containing Dy, Er and Yb possess a Kramers doublet ground state. The CEF scheme is somewhat different in LiGdF$_4$ and LiTmF$_4$. Owing to zero orbital degree of freedom in Gd$^{3+}$, the interaction with the CEF is absent and therefore we would expect this system to be a realization of a three-dimensional dipolar-coupled Heisenberg model at low temperatures. In LiTmF$_4$, the ground state is a singlet and only a Van Vleck magnetic moment is found. We therefore find an intriguing series of compounds which should provide a testing ground for the understanding of dipolar-coupled systems as well as experimental and theoretical insight into quantum phase transitions.

Generally, a strain on the lattice not only has a profound consequences on the CEF environment but also on the magnetic interactions. The interaction of lattice vibrations and magnetism of a $R$ ion is based on the magnetoelastic interaction which is proportional to the normal modes of vibration $u(\nu)$ of the anions that surround it and transform according to representation $\Gamma_\nu$ of the $\bar{4}$ group.\cite{lovesey-prb-2000} The single-ion magnetoelastic interaction can be described by,\cite{thalmeier-book-1991, lovesey-prb-2000}
\begin{equation}
\mathcal{H}_{\rm ME} = - \sum_\nu \zeta(\nu) u(\nu) \mathbf{P}(\nu)
\label{eq:me_lovesey}
\end{equation}
where $\zeta$ is the coupling constant. The rotation group with angular momentum $l=2$ contains five quadrupole operators that are second order in $\mathbf{J}_\alpha$. The quadrupole operators are taken to be $\mathbf{P}(\Gamma_1)=\mathbf{O}_2^0$, $\mathbf{P}(\Gamma_2)=\mathbf{O}_2^{\pm 2}$, and $\mathbf{P}(\Gamma_{3,4})=\mathbf{O}_2^{\pm 1}$. In order to establish the selection rules for the matrix elements of Eq.~\ref{eq:me_lovesey}, let us consider the $4f$ wavefunction transforming as $\Gamma_n$ and $\Gamma_m$. The non-vanishing matrix elements must contain the identity ($\Gamma_1$) in their direct product $\Gamma_n \times \Gamma_\nu \times \Gamma_m$ and this usually greatly restricts the phonon modes with the correct symmetry that need to be considered.

Inelastic neutron scattering is a powerful technique which probes both the energy difference and the wavefunctions of the states. Other spectroscopic techniques used to study \LRF\ compounds could only probe the transition energies. Neutron time-of-flight instruments with position sensitive detector arrays are able to sample large portions of $({\mathbf Q},\omega)$ space simultaneously, where we define ${\mathbf Q}$ and $\hbar\omega$ as the wavevector and energy transfer from the neutron to the sample, respectively. The measured energy spectrum exhibits resonance peaks which are associated with transitions between crystal-field levels. The neutron probes both the creation and destruction of crystal-field excitation. A finite lifetime broadens the resonance and can be detected if it exceeds the resolution of the spectrometer. The partial differential neutron scattering cross-section for crystal-field excitations can be expressed in the dipole approximation as,

\begin{multline}
\frac{k_i}{k_f}\left.\frac{{\rm d}^2 \sigma}{{\rm d} \Omega {\rm d} \omega}\right|_{n\rightarrow m} \propto
f^2(|{\mathbf Q}|)
\sum_\alpha \left( 1 - \hat{Q}_\alpha^2\right)
p_n|\langle \Gamma_m | {\mathbf J}_\alpha | \Gamma_n \rangle|^2 \\
\times \delta (\hbar\omega + E_n - E_m),
\label{eq:neutron_xs}
\end{multline}
where $k_i$ and $k_f$ are initial and final neutron wavevectors and the magnetic form factor is given by $f(|{\mathbf Q}|)$. The thermal population factor of the initial state $| \Gamma_n \rangle$ is given by $p_n$ and is defined as $\exp(-\beta E_n)/Z$ and $\beta=1/k_{\rm B}T$ for an energy level $E_n$ and partition function $Z$ at temperature $T$. The $\alpha$ component of the unit vector along the direction of ${\mathbf Q}$ is defined as $\hat{Q}_\alpha = Q_\alpha/|{\mathbf Q}|$.

The resonance peaks in an inelastic spectrum allow us to deduce the crystal-field splitting of the rare-earth ions while the scattering cross-sections of different transitions relate to the wavefunctions of the CEF states.
Spin and lattice fluctuations are amongst the common scattering processes which can also lead to modes being observed in spectra. For the systems examined in this paper, both the lattice vibrations and CEF excitations are in the 0--60\,meV range. Phonon scattering cross-section is proportional to $|{\mathbf Q}|^2$. Conversely, the crystal-field excitations decrease with $|{\mathbf Q}|$ due to the magnetic form-factor $f^2(|{\mathbf Q}|)$. The temperature dependence of the scattering cross-sections is also rather different. In the case of magnons and phonons, the intensity scales in accordance with Bose statistics, while the population of crystal-field levels obeys Boltzmann statistics, as given in Eq.~\ref{eq:neutron_xs}. Therefore, in most cases the $|{\mathbf Q}|$ and temperature dependence of excitations gives us the ability to relate them to the crystal field.

\section{Experimental details}





Powder and crystal samples of Li$R$F$_4$ were synthesized from the melt of LiF and $R$F$_3$ in glassy carbon crucibles. LiF was obtained from Li$_2$CO$_3$ (Alfa, 5N) and HF acid (Merck, suprapur, 40\%) followed by a HF gas treatment at 400$^\circ$C. The respective $R$F$_3$ were synthesized from the oxides (Metall Rare Earth Ltd., $R_2$O$_3$ 5N for $R$ = Gd, Ho, Er, Tb, and Yb$_2$O$_3$ 6N). The oxides were dissolved in HNO$_3$ (Merck, suprapur, 65\%), the fluorides precipitated by HF acid, the products dried, and treated with HF gas at 400$^\circ$C. Starting materials were handled in a dry box and used in LiF:$R$F$_3$ molar ratios of 53:47 for $R$ = Ho, Tm, and Tb, 55:45 for $R$ = Er, and 58:42 for $R$ = Gd. The mixtures were molten at 880$^\circ$C for $R$ = Gd, Ho, Tm, and Yb and at 910$^\circ$C for $R$ = Er. Crystals were obtained upon slow cooling by the Bridgman technique. For powder samples, crystal pieces were crushed in a mortar and LiF excess removed by washing with water. Product purity was checked by powder x-ray diffraction. It confirmed the Scheelite-type Li$R$F$_4$ phase and the absence of any extra lines, e.g., of $R$F$_3$ or LiF, for all samples.

The LiGdF$_4$ single-crystal has been grown in a home-made Czochralski furnace with conventional resistive heating, with a vacuum system allowing an ultimate pressure limit better than 10$^{-4}$ Pa. The growth process was carried out in a high-purity (5N) Ar atmosphere, starting from LiF and GdF$_3$ powders (nominal purity 5N, purified by HF processing to prevent OH$^{-}$ contamination, from AC Materials, Orlando, FL, USA) mixed in the ratio 68:32.\cite{chai-spie-1993} The growth has been carried with a sample rotation rate of 5\,rpm, a pulling rate 0.5\,mm/h, and melt temperature around 795$^{\circ}$C. Due to highly incongruent nature of the growth, the effective quantity grown is only $\approx$10\% of the available material. Using an x-ray Laue technique, the crystallographic axis of the boule has been identified and a mono crystalline sample was oriented and cut.

Inelastic neutron scattering measurements were performed using the time-of-flight spectrometers: FOCUS (PSI, Switzerland), TOFTOF (FRMII, Germany), LRMECS (IPNS, USA) and MERLIN (ISIS, UK).\cite{mesot-focus, Unruh-toftof, loong-lrmecs, bewley-merlin} Incident neutron energies in the range of 10-100\,meV were employed and data were collected at several temperatures between 4 and 300\,K. Magnetization measurements were performed on single-crystal samples of \LRF\ using a SQUID-based magnetometer (Quantum Design).

\section{Experimental results}

\subsection{LiHoF$_4$}

\begin{figure}
\centering
\includegraphics[width=\columnwidth,clip=]
{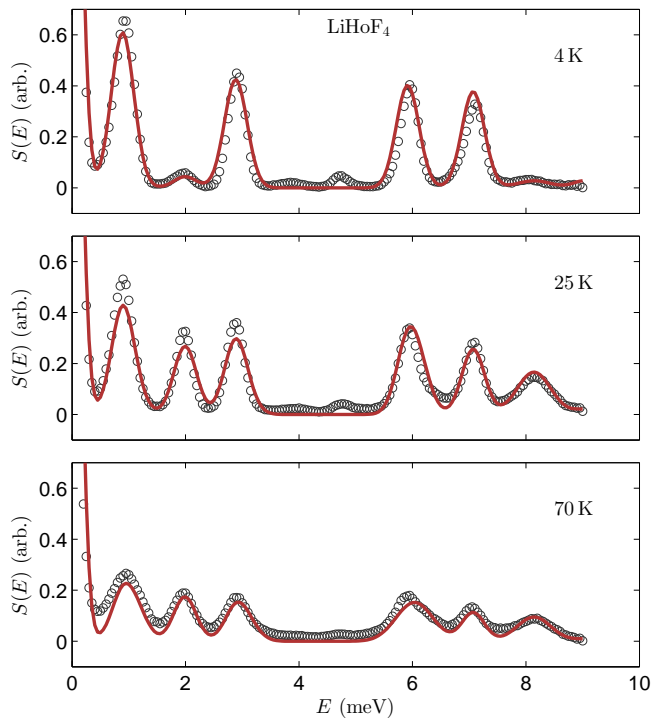}
\caption{(Color online). Neutron spectra recorded for LiHoF$_4$ at three different temperatures using TOFTOF spectrometer. The data were recorded for a 14.2\,g powder sample of LiHoF$_4$ using $E_i = 10.4$\,meV. The red line shows the calculated intensity based on the CEF parameters discussed in the text.
\label{fig:lihof4:cf}}
\end{figure}

The CEF parameters of LiHoF$_4$ have already been reported elsewhere based on magnetic susceptibility, Raman spectroscopy and EPR.\cite{hansen-prb-1975, gifeisman-1978, beauvillain-jmmm-1980, magarino-prb-1980, christensen-prb-1979a, gorller-book-1996,  salaun-condmatt-1997a, ronnow-prb-2007} However, no measurements using inelastic neutron scattering have been reported thus far. The $J=8$ manifold in LiHoF$_4$ is split into four doublet and nine singlet states by the CEF (see Eq.~\ref{eq:ions}) where the ground state is a mixture of levels $J_z = \pm 7, \mp 5, \pm 3, \mp 1$. Figure~\ref{fig:lihof4:cf} shows excitations collected at 4, 25 and 70\,K. Crystal-field excitations are clearly resolved in the 1 to 9\,meV energy transfer range. At low temperature excitations between the ground state to $n$th excited state dominate the spectra. As the temperature is increased high-order excited states become more populated leading to a redistribution of spectral weight. A weak mode is observed close to 4.7\,meV which disappears on warming the sample, the origin of this peak is unclear. The CEF parameters were deduced from diagonalizing the Hamiltonian defined in Eq.~\ref{eq:crystal_field} and fitting the inelastic neutron spectra for all three temperatures simultaneously. The fit parameters are in excellent agreement with those reported previously for LiHoF$_4$.\cite{hansen-prb-1975, gifeisman-1978, beauvillain-jmmm-1980, magarino-prb-1980, christensen-prb-1979a, gorller-book-1996,  salaun-condmatt-1997a, ronnow-prb-2007}

\begin{figure}
\includegraphics[width=\columnwidth]{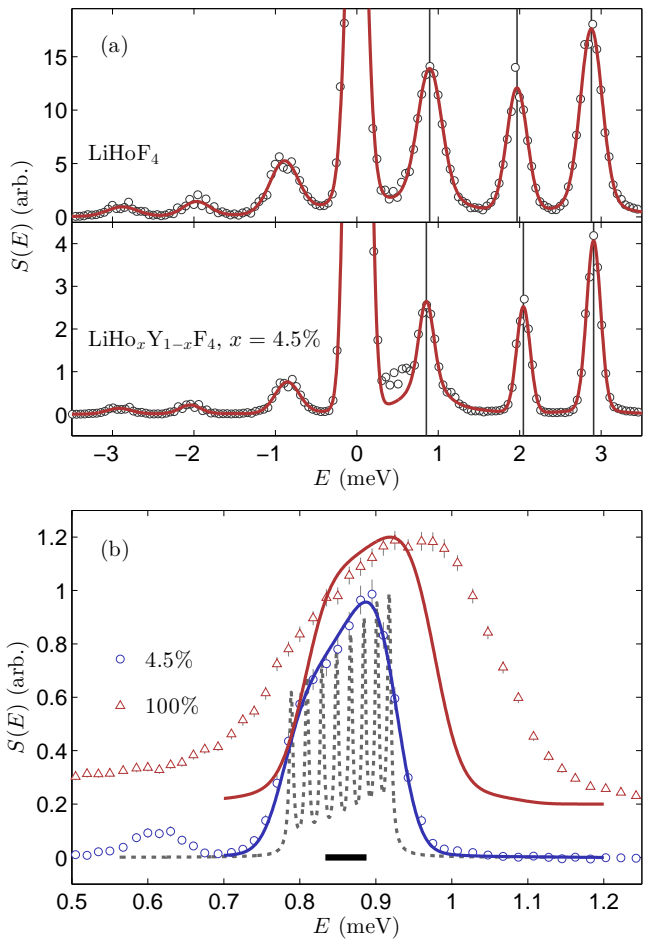}
\caption{(Color online) (a) Comparison of the time-of-flight spectra measured on TOFTOF for LiHoF$_4$ and LiHo$_x$Y$_{1-x}$F$_4$, $x = 0.045$ measured at 13\,K. The red line is a fit to the spectrum using Gaussians with black vertical lines indicating the CEF peak positions. (b) Higher-resolution spectrum of the ground state to first excited state transition for LiHoF$_4$ and LiHo$_x$Y$_{1-x}$F$_4$, $x = 0.045$, showing enhanced broadening of the pure compound relative to the diluted one. Data were recorded at 4\,K. The FWHM of the elastic line is shown by the black bar at the bottom. The dotted line shows RPA calculations discussed in the text.
\label{fig:lihof4:cf2}}
\end{figure}

Generally, it is assumed that dilution of LiHoF$_4$ by Y does not significantly influence the Ho CEF parameters.\cite{jensen-jpcm-1989} To verify this assumption, we have performed high-resolution inelastic neutron scattering measurements of powder samples with $x = 0.045$ and 1. Figure~\ref{fig:lihof4:cf2}(a) shows spectra collected on the energy gain and loss sides in the $-3.5 < E < 3.5$\,meV energy window at 13\,K. CEF excitations are observed close to 0.9, 2.0 and 2.9\,meV in both compounds. A small difference of less than 5\% is found in the transition energies between the pure and diluted systems. However, in this high-resolution configuration it was not possible to probe a sufficient number of transitions to allow for an accurate refinement of the complete set of CEF parameters.

To further compare the CEF excitations, the ground state to first excited state excitation was measured with higher resolution at 4\,K as shown in the inset of Fig.~\ref{fig:lihof4:cf2}(b). The line width of the pure compound is approximately five times broader than the instrumental resolution. For the LiHo$_x$Y$_{1-x}$F$_4$, $x = 0.045$, sample a splitting of the line is observed. The CEF mode is also much narrower than in LiHoF$_4$.

To simulate the dilute system, we employ random phase approximation (RPA) including only the CEF and hyperfine interactions; the results are shown by a dotted line in Fig.~\ref{fig:lihof4:cf2}(b). A small shift of $-0.037$\,meV of the correlation function was included to account for a slight modification of the CEF parameters in comparison with the pure system. The RPA calculation is then convoluted with the instrumental resolution (FWHM$ = 0.054$\,meV). We find that the overall splitting of the 0.9\,meV CEF mode of about 0.13\,meV comes mainly from the hyperfine splitting of the $\Gamma_{3,4}$ ground state (see Table~\ref{tab:energy_levels} for representations of CEF levels). Our simulations are found to very accurately reproduce both the measured CEF line width and asymmetry.

In the parent LiHoF$_4$ compound, dipole-dipole couplings become important. The dispersion of the first excited $\Gamma_2$ state remains quite small while the anisotropy splits the longitudinal and transverse modes by about 0.1\,meV. The RPA calculation folded with the instrumental resolution is shown in Fig.~\ref{fig:lihof4:cf2}(b). The dipole-dipole interaction increases the line width of the 0.9\,meV excitation compared to the dilute system. Although the line shape has a similar asymmetric form, the dispersion of the excitations is insufficient to account for the observed line width of 0.3\,meV. This would suggest that we require more than just dipolar interactions in order to explain the line width of the $\Gamma_{3,4}\rightarrow\Gamma_{2}$ transition in LiHoF$_4$. A plausible mechanism could be magnetoelastic coupling evidence for which is argued for in LiTmF$_4$ and LiYbF$_4$ later in this paper. Since Y and Ho ions are rather different in mass, the coupling strength between CEF and phonon modes could also be rather different in the two samples. An accurate knowledge of a possible modification of CEF environment by Y ions may be important in studies of, for example, LiHo$_x$Y$_{1-x}$F$_4$ systems where it has been assumed that the dilution of Ho ions does not alter the CEF environment.\cite{schechter-prb-2008, tabei-prb-2008, nikseresht-unpub} Further investigation is required to examine the origin of the CEF broadening.

\subsection{LiErF$_4$}

\begin{figure}
\centering
\includegraphics[width=\columnwidth,clip=]
{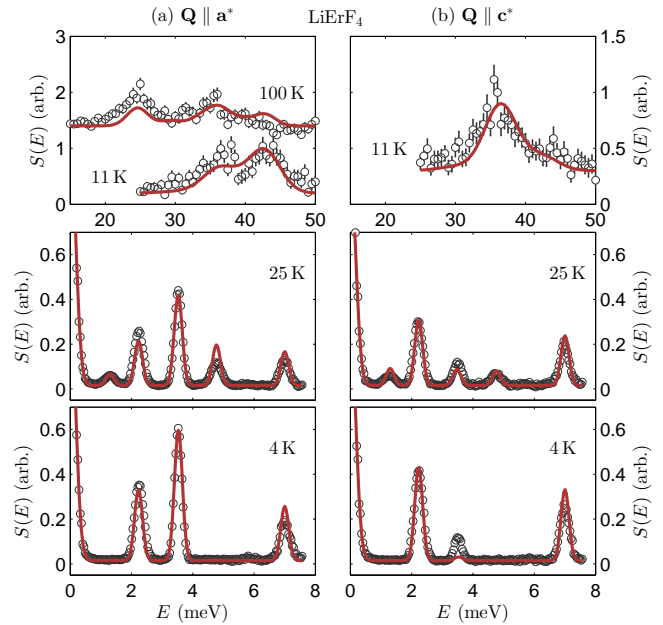}
\caption{(Color online). Neutron spectra recorded for a single-crystal of LiErF$_4$ at 4, 11, 25 and 100 K.
Panels and (a) and (c) show cuts taken with scattering wavevectors along $\mathbf a^\ast$ and $\mathbf c^\ast$, respectively. The red plots show results of simulated neutron scattering cross-sections based on the best-fit CEF parameters shown in Table~\ref{tab:cf_params}. Data for energy transfers less than 8\,meV were recorded on FOCUS ($E_i = 12$\,meV), while high-energy measurements up to 50\,meV were obtained using LRMECS ($E_i = 80$\,meV).
\label{fig:lierf4:cf}}
\end{figure}

The $J=15/2$ multiplet in LiErF$_4$ contains 8 doubly degenerate CEF levels. The present neutron scattering measurements were performed on a single-crystal sample such that the scattering wavevector was orientated along two directions: ${\mathbf a}^\ast$ and ${\mathbf c}^\ast$. This allows us to probe in greater detail the transition matrix elements due to the wavevector-dependent polarization factor in Eq.~\ref{eq:neutron_xs}. Data were collected at several temperatures between 4 and 100\,K. Thus, not only transitions from the ground state could be investigated, but also from higher-lying states which are populated at elevated temperatures. Figure~\ref{fig:lierf4:cf} shows the measured CEF excitations in LiErF$_4$. The first three transitions at approximately 2.2, 3.5, and 7.0\,meV can be easily resolved in measurements at 4\,K.
Transitions between electronic states can be well described by considering purely CEF transitions between the Kramers doublet states $\Gamma_{5,6}$ and $\Gamma_{7,8}$ which transform according to the irreducible representations of the $\bar{4}$ double group. Previous studies have demonstrated that the ground state possesses $\Gamma_{7,8}$ symmetry.\cite{hansen-prb-1975} Qualitatively one can observe that the 3.5\,meV mode is significantly stronger for $\Qb \parallel {\mathbf a}^\ast$ than $\Qb \parallel {\mathbf c}^\ast$. From this we can infer that the $\mathbf{J}_z$ matrix element is non-zero and that this excited state must belong to the same irreducible representation as the ground state. The same argument can be applied to the excitation at 43\,meV which transforms as $\Gamma_{7,8}$. Indeed, these observations match our calculated energy level scheme tabulated in Table~\ref{tab:energy_levels}.
The simultaneous fit of the neutron spectra is shown in Fig.~\ref{fig:lierf4:cf}. This refinement goes beyond that presented in Ref.~\onlinecite{kraemer-science-2012} by considering modes above 25\,meV which further constrain the CEF parameters. Based on our refined CEF parameters, we obtain a good agreement with the measured spectrum at all temperatures studied and for transitions measured up to 50\,meV. The 3.5\,meV excitation when measured with $\mathbf{Q}$ parallel to $\mathbf{c}^\ast$ appears to be significantly stronger than found in our model. It may be possible that the higher-energy CEF excitations hybridize with phonon modes which our model does not account for, however our current data cannot verify this. Another possible explanation is a small misalignment of the crystal.

\subsection{LiTmF$_4$}

\begin{figure}
\centering
\includegraphics[width=\columnwidth,clip=]
{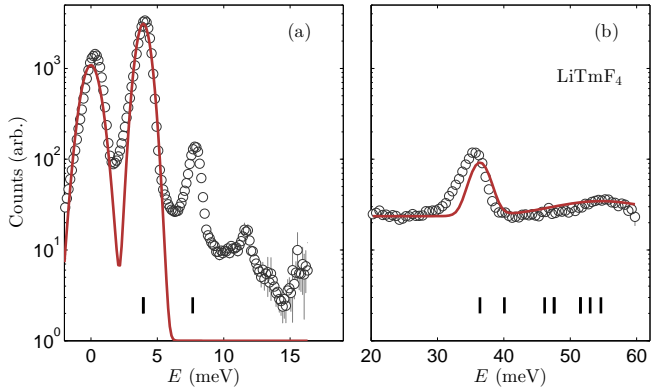}
\caption{(Color online) Inelastic neutron spectrum collected from a powder sample of LiTmF$_4$ at 5\,K collected using MERLIN spectrometer. Panels (a) and (b) show data collected using 20 and 80\,meV incident energy, respectively. Cuts were take centered on $|\Qb| = 1.4$\,\AA$^{-1}$ in panel (a) and 3\,\AA$^{-1}$ in panel (b). Calculated spectrum based on CEF parameters from Christensen\cite{christensen-prb-1979b} is plotted in red. Transition energies from ground state to excited states are shown by vertical black lines.
\label{fig:litmf4:cf}}
\end{figure}

In the $\bar{4}$ point symmetry CEF, the $(2J + 1) = 13$-fold degenerate electronic ground state of Tm$^{3+}$ splits into 7 singlet and 3 doublet states. Previous optical spectroscopy on LiTmF$_4$ has not been able to resolve the lowest lying CEF levels.\cite{jenssen-prb-1975, christensen-prb-1979b} To this end, we have used a powder sample of LiTmF$_4$ to examine the CEF excitations up to 60\,meV. Our results are shown in Fig.~\ref{fig:litmf4:cf}. Below 20\,meV, we find three, nearly equally spaced peaks centered on 3.96(2), 7.9(1) and 11.6(6)\,meV. We further find a strong CEF mode at 35.34(5)\,meV and a broad mode, which could be more than one transition, at 54.8(4)\,meV. The data do not allow for a reliable refinement of the CEF parameters. We therefore consider the refinement reported by Christensen\cite{christensen-prb-1979b}. The CEF modes are convoluted with Gaussians to approximate the effect of resolution. Using these parameters we expect a CEF transition from the ground state to a state at 7.66\,meV. However, our calculations show that the transition matrix between these two states is many orders of magnitude smaller than the 3.96\,meV excitation.

It is interesting to remark that the lowest three modes appear at multiples of about 3.96\,meV. Furthermore, the 3.96\,meV mode is very strong, significantly stronger than the incoherent scattering. A successive decrease by around an order of magnitude in intensity is found for the 7.9 and 11.6\,meV excitations. When a neutron traverses the sample, the most dominant scattering process is typically where a scattered neutron is transmitted through the sample to the detector. However, this need not be the case as in thick or strongly scattering samples, a neutron can undergo several scattering events before leaving the sample. We conclude that the 7.9 and 11.6\,meV peaks are due to such multiple scattering.

\begin{figure}
\centering
\includegraphics[width=\columnwidth,clip=]
{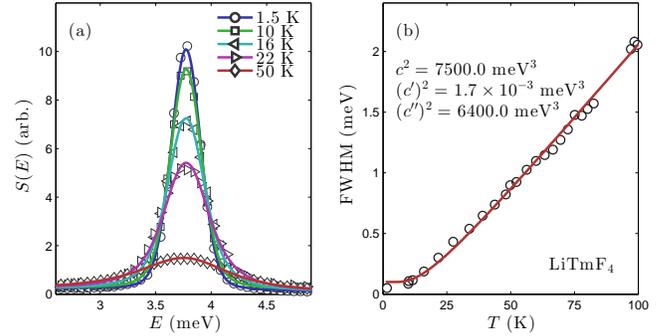}
\caption{(Color online) (a) Neutron scattering spectrum of LiTmF$_4$ powder showing temperature dependence of the CEF excitation at 3.77\,meV. Measurements were obtained using FOCUS with $E_i = 7$\,meV. The spectra were fit by a Lorentzian function. (b) Line width extracted from fitting the CEF excitation at temperatures from 1.5 to 100\,K. The solid line is a simulation of the CEF lineshape broadening by a magnetoelastic model described in the text.
\label{fig:litmf4:broadening}}
\end{figure}

Unlike the other CEF excitations presented in this paper, we discovered that in LiTmF$_4$ there is a change in line width as a function of temperature of the mode corresponding to excitation between the ground and first excited CEF states. Figure~\ref{fig:litmf4:broadening} shows how the CEF transition decreases in amplitude and at the same time becomes significantly broader from 1.5 to 100\,K. No shift in position of the peak is observed. Figure~\ref{fig:litmf4:broadening}(b) shows that the broadening has an initial upturn around 10\,K and is approximately linear at higher temperatures. A similar effect was found from near-infrared spectroscopy where the transition between ground state and an excited state at 2.6\,eV was found to broaden in the 5--250\,K temperature range.\cite{christensen-prb-1979b} Generally, CEF states are subject to interaction with phonons, spin fluctuations and charge carriers which impact the lifetime of the CEF states. In metallic rare-earth compounds where charge carriers are dominant in the relaxation mechanism, the line width of CEF transitions can scale linearly with temperature.\cite{korringa-physica-1950} However, this is unlikely to be the case for LiTmF$_4$ which is a good insulator. Magnetic ions with unfilled $3d$ and $4f$ shells are often found in different valence states. Such states can exhibit inter-configurational or valence fluctuations.\cite{luthi-book-2005} For compounds consisting of Ce, Sm, Eu, Tm and Yb the energy separation between two integral valence states can be 0 to 2\,eV. These fluctuations can broaden the CEF transitions, however this effect is mediated by conduction electrons and so in LiTmF$_4$ (an insulator) seems very unlikely. Instead, we consider the role of coupling of lattice vibrations to CEF states which can lead to broadening of CEF lines as function of temperature.

Let us consider a simple magnetoelastic model where lattice vibrations contribute to the relaxation of a transition between CEF energy levels. This effect has been previously discussed when lanthanide ions are substituted for Y in YBa$_2$Cu$_3$O$_{7-\delta}$.\cite{lovesey-prb-2000, boothroyd-prb-2001} We consider the effect of acoustic phonons on a transition between $|a\rangle$ and $|b\rangle$ states with all other intermediate levels $|\gamma\rangle$ higher in energy than $|b\rangle$. The energy of the $|\gamma\rangle$ level is $E_\gamma$ relative to the ground state $|a\rangle$. This leads to the linewidth $\sigma_{\rm ME}$ in the magnetoelastic model to be expressed as the summation over all the phonon modes of symmetry $\Gamma_\nu$,\cite{lovesey-prb-2000, boothroyd-prb-2001}
\begin{equation}
\sigma_{\rm ME} = \sum_\nu\sigma_\nu,
\end{equation}
and,
\begin{align}
\sigma_\nu
 &=& c_\nu^2\left|\langle a|\mathbf{P}(\nu)|b\rangle\right|^2\frac{Z_\nu(E_b)}{E_b}\coth(\beta E_b/2) \nonumber\\
 &+& \sum_\gamma (c'_\nu)^2\left|\langle a|\mathbf{P}(\nu)|\gamma\rangle\right|^2
\frac{Z_\nu(E_\gamma)}{E_\gamma}n(E_\gamma) \nonumber\\
 &+& \sum_\gamma(c''_\nu)^2\left|\langle b|\mathbf{P}(\nu)|\gamma\rangle\right|^2
\frac{Z_\nu(E_\gamma-E_b)}{E_\gamma-E_b}n(E_\gamma-E_b).
\label{eq:litmf4_mewidth}
\end{align}
$Z_{\nu}(E)$ is the density of phonon modes of symmetry $\nu$ and $n(E) = 1/(e^{\beta E}-1)$. The parameters $c_{\nu}$, $c'_{\nu}$ and $c''_{\nu}$ are proportional to the magnetoelastic coupling constants and depend upon the symmetry of the modes. We shall make the simplification of assuming the coupling constants are independent of $\nu$. In order to proceed to simulate the relaxation of the CEF mode as a function of temperature, we shall assume a simple Debye density of states model with Debye temperature $\theta_{\rm D} = 279$\,K or 24.0\,meV which is also independent of $\nu$. The value of the Debye temperature is taken in the first approximation to be same as in LiYF$_4$.\cite{blanchfield-jpoc-1979} Although we could include all $2J + 1$ levels in our calculation, only the first three CEF levels are below $\theta_{\rm D}$ and therefore in our model higher CEF do not play a role in the broadening mechanism. We employ CEF parameters given by Christensen \emph{et al.}\cite{christensen-prb-1979b} and shown in Table~\ref{tab:cf_params}. By allowing $c$, $c'$ and $c''$ to vary, we can obtain a reasonably good agreement with the experimental results as plotted in Fig.~\ref{fig:litmf4:broadening}(b). Our model has the same initial slow upturn until around 10\,meV with a linear dependence on temperature above when $E_b, E_\gamma \ll T$. We find that despite including all possible quadrupole operators ($m = 0,\pm1,\pm2$), the dominant contribution to the line broadening comes from phonons with $\Gamma_{3,4}$ symmetries. Although the magnetoelastic coupling simulations we present here are rather simplistic, we show that the relaxation rate of CEF levels in LiTmF$_4$ is likely to be related to the CEF distortion created by lattice vibrations.

\subsection{LiYbF$_4$}
\label{sec:liybf4}

\begin{figure}
\centering
\includegraphics[width=\columnwidth,clip=]
{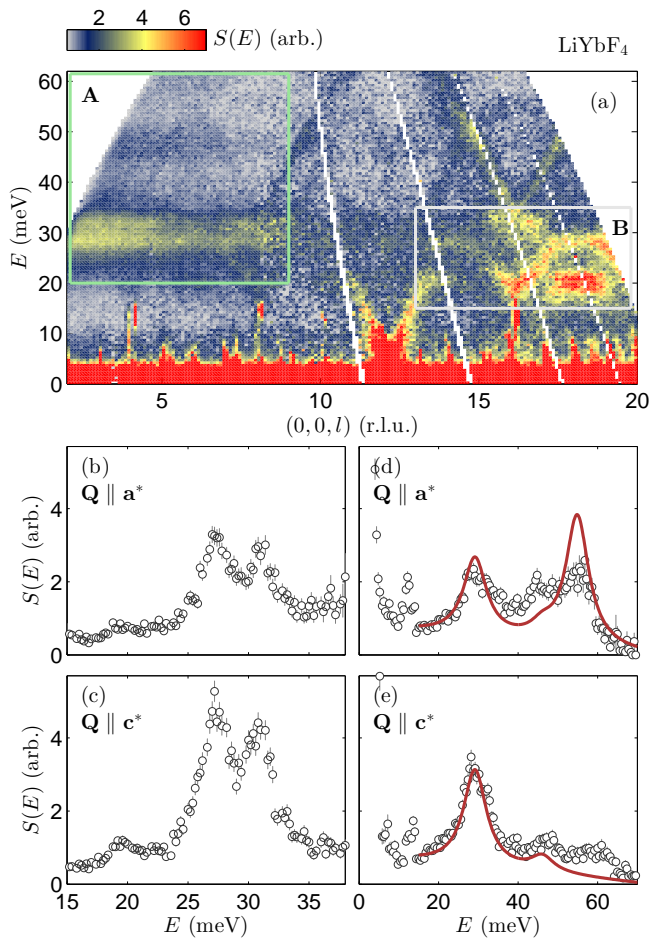}
\caption{(Color online) Inelastic neutron scattering measurements on a single-crystal sample of LiYbF$_4$ at 10\,K using MERLIN.
Panel (a) shows a slice through the spectra collected along $(0,0,l)$. The regions A and B are discussed in the text.
Panels (b) and (c) show data collected using $E_i = 46$\,meV cut along directions parallel to $a$ and $c$ axes, respectively. Panels (d) and (e) show CEF modes between 20 and 65\,meV measured using 92\,meV incident neutrons. The red line shows CEF parameter refinement discussed in the text.
\label{fig:liybf4:cf}}
\end{figure}

No neutron scattering measurements have been reported on LiYbF$_4$ so far. The CEF splitting was investigated by optical studies.\cite{salaun-condmatt-1997b, miller-jap-1970, dong-pss-2003} We have performed inelastic neutron scattering measurements on a single-crystal sample of LiYbF$_4$ at 10\,K with crystallographic ${\mathbf a}^\ast$ and ${\mathbf c}^\ast$ axes aligned in the horizontal scattering plane. The MERLIN spectrometer was employed for this experiment. A primary incident neutron energy of 92\,meV with Fermi chopper spinning at 500\,Hz was used with the instrument running in multi-repetition mode such that a secondary 46\,meV incident energy was also accessible. The $J=7/2$ ground state multiplet of LiYbF$_4$, is split in $\bar{4}$ symmetry into four doubly-degenerate CEF levels. According to previous optical measurements the excited states are located between 30 and 55\,meV above the ground state.\cite{miller-jap-1970} All of these levels are therefore captured in our spectra as at low temperatures we do not expect thermal population of the first excited state at approximately 30\,meV. To refine the CEF parameters we have performed measurements with momentum transfer along ${\mathbf a}^\ast$ and ${\mathbf c}^\ast$.

A representative slice through reciprocal space showing excitations collected at 10\,K is shown in Fig.~\ref{fig:liybf4:cf}(a). At small momentum transfer we find distinct flat modes around 30 and 55\,meV, highlighted in region A of Fig.~\ref{fig:liybf4:cf}(a). These clearly decrease at larger $|\Qb|$ as expected for a magnetic excitation. Conversely, at larger $|\Qb|$ we observe an increase of scattering from phonons. Since Yb is the heaviest ion in the LiYbF$_4$ compound, we expect the lowest energy modes to be dominated by lattice vibrations from Yb and higher-energy branches to be due to vibrations of Li and F ions. In region B of Fig.~\ref{fig:liybf4:cf}(a), we find strong scattering from phonon modes between 20 and 30\,meV. There appears to be a flattening of the mode close to 30\,meV which may be indicative of mixing of electronic and phonon degrees of freedom.

For the magnetic modes visible in region A of Fig.~\ref{fig:liybf4:cf}(a), rather unexpectedly we find that the mode around 28.9\,meV can clearly be resolved into two excitations centered at 27.2(1) and 30.9(2)\,meV as shown in Fig.~\ref{fig:liybf4:cf}(b) and (c). Using 92\,meV incident energy, we observe three modes at approximately 28.9(1), 46.2(5) and 55.2(5)\,meV, as shown in Fig.~\ref{fig:liybf4:cf}(d) and (e). These levels are in good agreement with previous optical measurements.\cite{miller-jap-1970} The excitations decrease in amplitude at higher $|\Qb|$ and at higher temperatures (100-300\,K, not shown here) which is indicative of magnetic origin.

The CEF parameters shown in Table~\ref{tab:cf_params} have been extracted by fitting the neutron spectra along the two perpendicular directions as well as the magnetization data. To reduce the number of free parameters, fitting was done assuming $\bar{4}2m$ site symmetry such that $B^0_6 = B^4_6({\rm s})=0$. Allowing these parameters to vary did not produce a statistically better fit to the data. The extraction of the CEF parameters was found not to be strongly influenced by the neutron intensity. However, the CEF transition energies as well as magnetization data were found to be very sensitive on the values of the CEF parameters. A large phonon-scattering contribution above 100\,K did not allow for an accurate extraction of the CEF intensities as a function of temperature and was not used in the fit.

Comparing Fig.~\ref{fig:liybf4:cf}(d) and (e), we find a strong peak in $\mathbf{Q} \parallel {\mathbf a}^\ast$ but not in $\mathbf{Q} \parallel {\mathbf c}^\ast$ which implies that the ground state and highest excited state at 55.2\,meV are described by the same irreducible representation $\Gamma_{5,6}$. This leaves the levels at 29.4 and 44.2\,meV to be of $\Gamma_{7,8}$ symmetry.

\begin{figure}
\centering
\includegraphics[width=\columnwidth,clip=]
{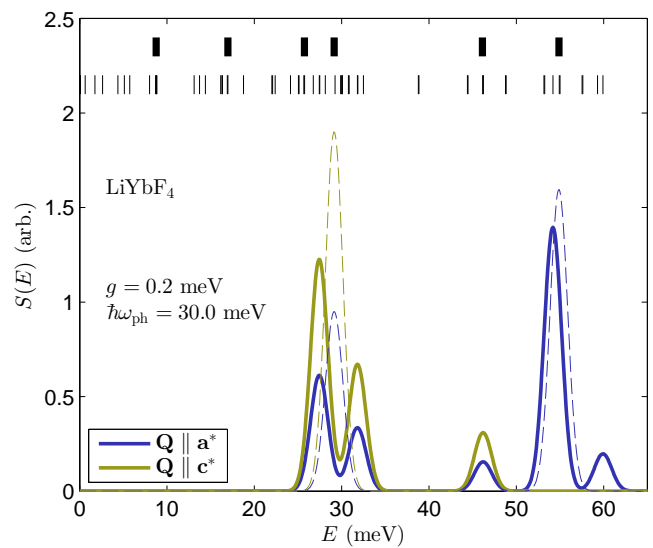}
\caption{(Color online). Simulation of the influence of magnetoelastic coupling on the magnetic spectrum of LiYbF$_4$. The dashed plots indicate the unperturbed CEF peak centers. The level scheme for the unperturbed ($g=0$) and perturbed ($g=0.2$\,meV) states by bold and thin lines, respectively.
\label{fig:liybf4:me_coupling}}
\end{figure}

The splitting of the 28.9\,meV mode is surprising and is reminiscent of magnetoelastic coupling. However, this would require a flat phonon branch with the right symmetry over an extended part of the Brillouin zone.\cite{thalmeier-prl-1982, jensen-prb-2007, adroja-prl-2012} From infrared spectroscopy, this appears to be plausible as several longitudinal and transverse optical modes are found close to 30\,meV.\cite{salaun-condmatt-1997a} In order to determine to what extent this interaction plays a role in influencing the magnetic spectrum we shall consider a simple model that includes the following terms in the Hamiltonian,\cite{boothroyd-prl-2001}
\begin{equation}
\mathcal{H} = \mathcal{H}_{\rm CEF} +
\sum_\nu \hbar\omega_{\rm ph}a_\nu^\dag a_\nu +
g(a_\nu + a_\nu^\dag)\mathbf{P}(\nu),
\label{eq:me_coupling}
\end{equation}
where $a$ and $a^\dag$ are phonon annihilation and creation operators for distortions with vibrational frequency of $\hbar\omega_{\rm ph}$ and $g$ is the constant which describes the strength with which CEF modes couple to phonons. If the $4f$ wavefunction transforms as $\Gamma_n$ and $\Gamma_m$ and the representation $\Gamma_\nu$ is the symmetry of the quadrupole operator $\mathbf{P}$, then the direct product $\Gamma_n\times\Gamma_\nu\times\Gamma_m$ must contain the identity in order for coupling to be allowed. For transition from ground state to first excited state in LiYbF$_4$, the wavefunction transform as $\Gamma_{5,6}$ and $\Gamma_{7,8}$ (see Table~\ref{tab:energy_levels}), respectively. We therefore should include in our calculation the operators $\mathbf{P}(\Gamma_2)$ which have the correct symmetry. The Hamiltonian in Eq.~\ref{eq:me_coupling} can be readily diagonalized in the basis of $J_z$ and phonon states.

Figure~\ref{fig:liybf4:me_coupling} shows the calculation of the magnetic part of the neutron cross section from the matrix element of $\mathbf{J}$, given by Eq.~\ref{eq:neutron_xs}. For the case when $g=0$, the CEF modes do not couple to the dynamic lattice distortions and the magnetic spectrum is unaffected. Let us now consider double-degenerate local distortion of $\Gamma_2$ symmetry with energy of $\hbar\omega_{\rm ph} = 30$\,meV. This creates vibronic states which for a small coupling constant of $g=0.2$\,meV transfers intensity into phonon-like states. From our calculations, shown in Fig.~\ref{fig:liybf4:me_coupling}, we find that this leads to a splitting of the 28.9\,meV mode. This bears strong resemblance to our data in Figs.~\ref{fig:liybf4:cf}(b) and \ref{fig:liybf4:cf}(c). We further note that as scattering from phonons increases at larger $|\Qb|$, we observe what appears to be a flattening of an otherwise dispersing phonon mode at the CEF level energy [Fig.~\ref{fig:liybf4:cf}(a), region B]. This would support our notion that CEF are perturbed by phonons but we must note that our single-frequency phonon mode approach is somewhat limited and cannot be expected to hold exactly. In reality we would expect some dispersion of phonon modes and the broadness of the transitions measured would support this. Nevertheless, we believe our magnetoelastic model gives qualitative agreement with our observations. We finally note that a similar effect has been observed in Raman scattering measurements in LiTbF$_4$ where a partial hybridization of the electronic and phonon wavefunctions has been shown to lead to a transfer of intensity between phonon and electronic components.\cite{doerfler-zpb-1985}

Another possibility is that the tripositive Yb ion is not entirely stable -- it may fluctuate between being Yb$^{3+}$ and Yb$^{2+}$ (with a filled $4f$ shell). Valence properties of Yb ions have been reported under pressure in Ref.~\onlinecite{ylvisaker-prl-2009}. However, it is not evident why valence fluctuations should lead to a splitting of the Kramer's doublet, and, in addition F ions should be quite effective in stabilizing the tripositive state of the Yb ions. Furthermore, we find that the higher levels at 45 and 55\,meV are significantly broader than the instrument resolution. From our measurements we cannot resolve the modes more clearly, nor determine whether it is the lifetime of the excited state which is responsible for the broadening or a physical effect which splits the levels. Whether or not it is the ground state, rather than the first excited state that is split remains an open question and should be addressed in further studies. The alternative scenario would be that there is an additional inequivalent Yb site in the crystal. However, our neutron powder diffraction measurements did not find any evidence for this.\cite{babkevich} In the absence of magnetic order which breaks time-reversal symmetry, it is not evident what would be responsible for lifting the ground state degeneracy other than a hybridization of a phonon branch with CEF excitations.

The dominant CEF parameters, listed in Table~\ref{tab:cf_params}, are significantly larger in LiYbF$_4$ compared to for example LiErF$_4$. The $B^0_2$ CEF parameter plays an important role in the splitting of the levels and to a first approximation is related to the high-temperature susceptibility as $1/\chi_c - 1/\chi_a \propto B^0_2$.\cite{hansen-prb-1975} Therefore, although our simulations do not reproduce the intensity of neutron scattering from the CEF excitations of LiYbF$_4$ particularly accurately, the neutron energy transfer at which the excitations are found and the susceptibility measurements provide a good restriction on the possible values of $B^0_2$. Since the critical transverse field also scales with $B^0_2$ we expect that our calculations should reproduce the phase diagram of LiYbF$_4$, described in \S~\ref{sec:phase_diag}, reasonably well.

\subsection{Summary of results}

\begingroup
\squeezetable
\begin{table}
\centering
\begin{tabular}{ l c c c c c c r}
\hline
\hline
\LRF    & $10^3B^0_2$   & $10^3B^0_4$    & $10^3B^4_4$(c)
        & $10^6B^0_6$   & $10^6B^4_6$(c) & $10^6|B^4_6$(s)$|$   & Ref.\\
\hline
Ho      & \textbf{-57.9}    & \textbf{0.309}      & \textbf{3.51}
        & \textbf{0.540}      & \textbf{63.1}     & \textbf{17.1}         &  $\ast$  \\
        & (3.2)     & (0.047)   & (0.60)
        & (0.14)    & (12)      & (3.3)          &  $\ast$  \\
        & -64.8	    & 0.426	    & 4.54	
        & 0.100	    & 85.6	    & 16.9          & \onlinecite{hansen-prb-1975}\\
        & -52.5	    & 0.280	    & 3.70	
        & 0.725	    & 70.44	    & 0.00           & \onlinecite{gifeisman-1978}\\
        & -52.2	    & 0.323	    & 3.59	
        & 0.522	    & 68.5	    & 0.00           & \onlinecite{christensen-prb-1979a}\\
        & -73.6     & 0.478	    & 4.69	
        & 0.100	    & 86.1  	& 11.8           & \onlinecite{beauvillain-jmmm-1980}\\
        & -56.2 	& 0.325	    & 3.61	
        & 0.181	    & 75.8	    & 0.00           & \onlinecite{gorller-book-1996}\\
        & -60.0     & 0.350     & 3.60
        & 0.400     & 70.0      & 9.80           & \onlinecite{ronnow-prb-2007}    \\
\hline
Er      & \textbf{58.1}     & \textbf{-0.536}     &\textbf{-5.53}
        &\textbf{-0.00625}    & \textbf{-106}   &\textbf{23.8}          & $\ast$ \\
        &(3.4)      & (0.032)   & (0.31)
        & (0.00041) & (6.1)     & (1.5)          & $\ast$ \\
        & 67.8 	    &-0.678	    & -6.83	
        &-0.080	    & -133   	& 24.3           & \onlinecite{hansen-prb-1975}\\
        & 49.2	    &-0.390	    & -4.14	
        &-0.899	    & -92.6	    &0.00            & \onlinecite{christensen-prb-1979a}\\
        & 76.3	    &-0.568	    & -6.37	
        & -1.72	    &-132	    & 23.4           & \onlinecite{beauvillain-jmmm-1980}\\
        & 55.5 	    &-0.565	    &-5.76	
        &-2.15	    &-111   	& 0.00           & \onlinecite{gorller-book-1996}\\
        & 47.8      & -0.53     & -5.39
        & -0.961    & -120      &0.00            & \onlinecite{heyde-jcs-1998}\\
        & 60.2 	    &-0.120	    &-4.33	
        & -1.90	    &-85.0	    & 22.7           & \onlinecite{kraemer-science-2012}\\
\hline
Tm      & 225	    &-1.54	    &-17.9	
        & 7.52	    & 307   	& 0.00           & \onlinecite{jenssen-prb-1975}\\
        & 230	    &-1.81	    &-19.4	
        & 2.78	    & 300	    & 57.6           & \onlinecite{christensen-prb-1979b}\\
        & 218   	&-1.62	    &-18.3 	
        & 7.91	    & 313	    & 0.00           & \onlinecite{gorller-book-1996}\\
        & 231       &-1.82      &-17.9
        & 2.82	    & 302	    & 0.0236         & \onlinecite{abubakirov-jpcm-2008}\\
\hline
Yb      &  \textbf{457}   & \textbf{7.75}      & \textbf{196}
        &  \textbf{0}        & \textbf{-9780}  & \textbf{0}              & $\ast$ \\
        &  (5.2)    & (0.12)    & (0.65)
        &  (0)      & (9.4)     & (0)            & $\ast$ \\
        & 737	    & 16.5	    & 176	
        & -18.4	    & -4070	    & 0.00           & \onlinecite{dong-pss-2003}\\
        & 720	    & 16.4	    & 177	
        & -18.4	    & -5150	    & 0.00           & \onlinecite{feng-2009}\\
\hline
\hline
\end{tabular}
\caption{CEF parameters $B_l^m$ in units of meV obtained in this work for \LRF\ compared to previous studies. Our refinements, with estimated uncertainties shown in brackets in the line below, are denoted by $\ast$. Values of CEF parameters from references where transformed into the coordinate system defined in text. The sign of $B^4_6$(s), which depends on the choice of the F ion coordination, is left undetermined.
\label{tab:cf_params}}
\end{table}
\endgroup

\begin{table}
\centering
\begin{tabular}{ l r l r l r l r}
\hline
\hline
& LiHoF$_4$ & & LiErF$_4$ & & LiTmF$_4$ & & LiYbF$_4$ \\
\hline
$\Gamma_{3,4}$  & 0.00  & \quad $\Gamma_{7,8}$  & 0.00  & \quad$\Gamma_{2}$      & 0.00  & \quad$\Gamma_{5,6}$ & 0.00 \\
$\Gamma_{2}$    & 0.89  & \quad $\Gamma_{5,6}$  & 2.23  & \quad$\Gamma_{3,4}$    & 3.94  & \quad$\Gamma_{7,8}$ & 29.14 \\
$\Gamma_{2}$    & 2.88  & \quad $\Gamma_{7,8}$  & 3.53  & \quad$\Gamma_{1}$      & 7.66  & \quad$\Gamma_{7,8}$ & 46.11 \\
$\Gamma_{1}$    & 5.91  & \quad $\Gamma_{5,6}$  & 7.00  & \quad$\Gamma_{2}$      & 36.40 & \quad$\Gamma_{5,6}$ & 54.88 \\
$\Gamma_{1}$    & 7.07  & \quad $\Gamma_{5,6}$  & 31.62 & \quad$\Gamma_{2}$      & 40.07 &                &  \\
$\Gamma_{3,4}$  & 9.03  & \quad $\Gamma_{5,6}$  & 36.39 & \quad$\Gamma_{1}$      & 46.14 &                &  \\
$\Gamma_{1}$    & 25.96 & \quad $\Gamma_{7,8}$  & 39.79 & \quad$\Gamma_{3,4}$    & 47.56 &                &  \\
$\Gamma_{1}$    & 32.69 & \quad $\Gamma_{7,8}$  & 42.85 & \quad$\Gamma_{3,4}$    & 51.54 &                &  \\
$\Gamma_{3,4}$  & 32.71 &               &       & \quad$\Gamma_{1}$      & 52.99 &                &  \\
$\Gamma_{2}$    & 33.74 &               &       & \quad$\Gamma_{2}$      & 54.61 &                &  \\
$\Gamma_{1}$    & 35.82 &               &       &                   &       &                &  \\
$\Gamma_{3,4}$  & 36.42 &               &       &                   &       &                &  \\
$\Gamma_{2}$    & 38.32 &               &       &                   &       &                &  \\
\hline
\hline
\end{tabular}
\caption{CEF level energies in meV relative to the ground state. The results were obtained from diagonalizing the CEF Hamiltonian using parameters given in Table~\ref{tab:cf_params}. The irreducible representation is given for each level where double subscripts label doubly degenerate levels.
\label{tab:energy_levels}}
\end{table}

The CEF parameters $B_l^m$ of \LRF\ ($R = $ Ho, Er, Tm and Yb) are listed in Table~\ref{tab:cf_params}. We note that our refinement is somewhat different from studies using infrared spectroscopy. This is because for the energy transfer range which we employed in our inelastic neutron scattering we can only excite transitions within the lowest $J$-multiplet. However, these are the most important energy levels and states in predicting the low temperature magnetism in which we are interested and therefore their accurate refinement is important. Our results are broadly in good agreement with previous studies.
A discrepancy is found for the less well known LiYbF$_4$.\cite{feng-2009,dong-pss-2003} Although the energy-level scheme is found to be in agreement with those measured, we cannot capture very well the \Qb-dependence of the intensities in our neutron scattering measurements. Our results in \S~\ref{sec:liybf4} imply hybridization between CEF and phonon modes which affects the refinement of CEF parameters.

\begin{figure*}
\centering
\includegraphics[width=0.8\textwidth,clip=]
{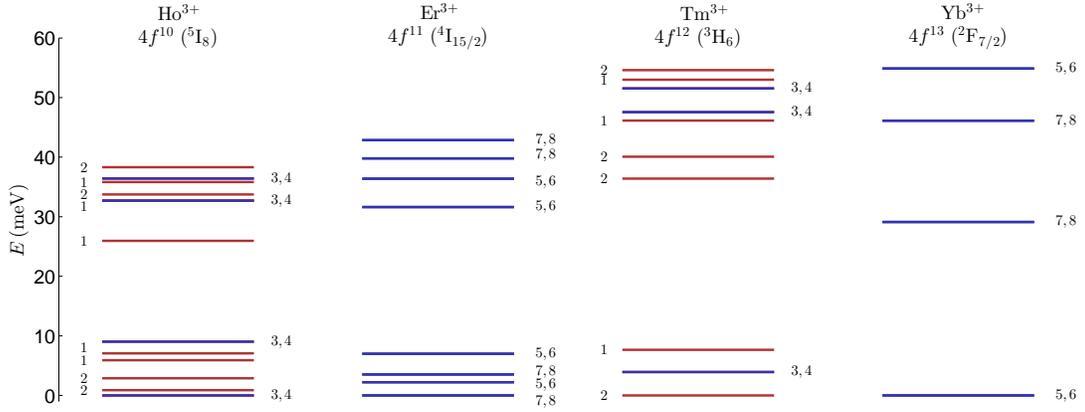}
\caption{(Color online). Splitting of the ground state multiplet $J$-sublevels of $R$ ions by the CEF as calculated from inelastic neutron scattering measurements. The labels next to the levels denote the symmetry of that level. Double indices indicate that the level is doubly degenerate. Levels plotted in red and blue denote singlet and doublet states, respectively.
\label{fig:liref4:cf}}
\end{figure*}

From our analysis we can diagonalize the CEF Hamiltonian to obtain the CEF level scheme shown in Fig.~\ref{fig:liref4:cf} for each compound. The CEF modes are found below 60\,meV. Based on the wavefunctions, we can determine the representation of each CEF state, summarized in Table~\ref{tab:energy_levels}. The symmetries agree perfectly with those reported for LiHoF$_4$, LiErF$_4$ and LiTmF$_4$.\cite{hansen-prb-1975, salaun-condmatt-1997a} However, ground and first excited states in LiYbF$_4$ in our work are opposite representations.\cite{salaun-condmatt-1997a} As expected, for LiErF$_4$ and LiYbF$_4$ all levels are doubly degenerate Kramer states. The ground state in LiErF$_4$ possess $\Gamma_{7,8}$ symmetry while in LiYbF$_4$ we find $\Gamma_{5,6}$ as the ground state. In the case of LiHoF$_4$, the ground state $\Gamma_{3,4}$ is a doublet with a singlet state $\Gamma_2$ directly above. The situation in LiTmF$_4$ is reversed as the ground state $\Gamma_2$ is a singlet but the first excited state is degenerate and described by representation $\Gamma_{3,4}$. This change causes a dramatic difference on the dipolar interactions between the two systems where the singlet ground state does not permit long-ranged ordering of ions in LiTmF$_4$ while LiHoF$_4$ realizes a dipolar-coupled ferromagnet below 1.53\,K.

\begin{figure}
\centering
\includegraphics[width=\columnwidth,clip=]
{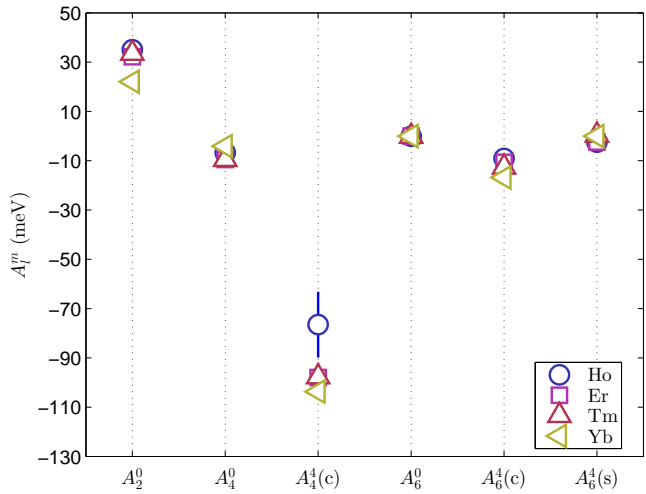}
\caption{(Color online). Extraction of the ion-independent $A^m_l$ parameter from the $B^m_l$ CEF parameters discussed in the text.
\label{fig:liref4:cfAlm}}
\end{figure}

The CEF Hamiltonian defined in Eq.~\ref{eq:total_Ham_simple} can be expressed in terms of the intrinsic CEF parameter $A^m_l$, where $A^m_l = B^m_l /(\theta_l \langle r^l \rangle)$. The CEF Hamiltonian is more commonly defined using $B^m_l$ parameters as the derivation of $\langle r^l \rangle$ is non-trivial. To obtain the ion-independent $A^m_l$ parameters we consider Dirac-Fock values of $\langle r^l \rangle$ tabulated by Freeman and Desclaux~\cite{freeman-jmmm-1979}. The values can then be extrapolated to deduce the CEF level scheme in other rare-earth \LRF\ systems using appropriate Stevens factors and radial wavefunctions for a given ion. Figure~\ref{fig:liref4:cfAlm} summarizes the CEF parameters obtained in our work converted to $A^m_l$. The $A^m_l$ parameters are consistent for all four ions considered.

\section{Phase diagram calculations}
\label{sec:phase_diag}

\begin{figure*}
\centering
\includegraphics[width=1.9\columnwidth,clip=]
{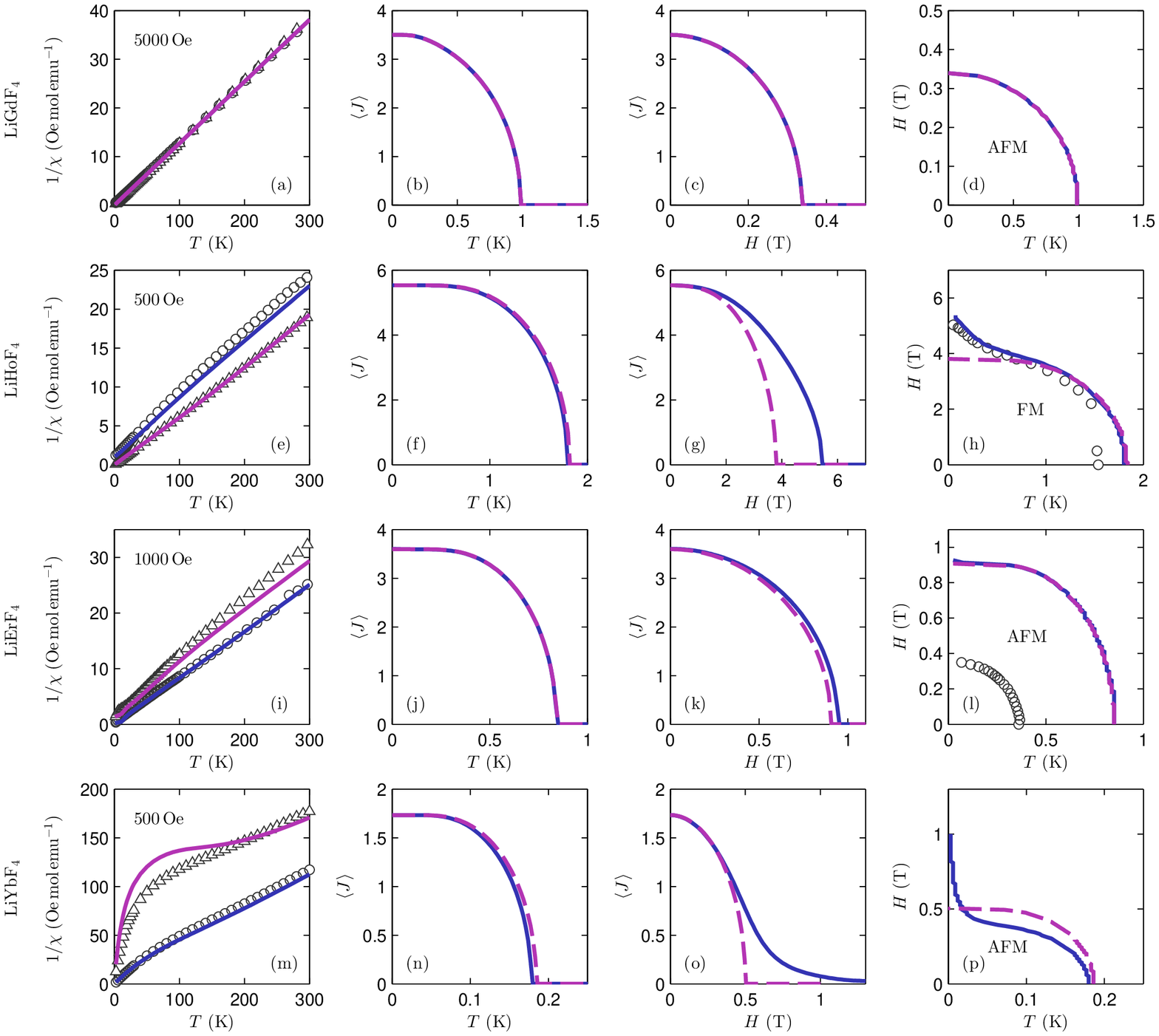}
\caption{(Color online). Overview of experimental and theoretical results of magnetization dependence on temperature and field for \LRF = Gd, Ho, Er and Yb. The left-most panels show the measured and calculated inverse-susceptibility in an applied field 500-5000\,Oe. Triangles (circles) denote the moment measured with applied field along the $a$ ($c$) axis. The center-left and right panels show the temperature and field dependence of the order parameter for each compound, respectively. The dashed (solid) line indicates calculations which (do not) include the effect of the hyperfine interaction. The right-most panels are calculations of the low-temperature magnetic phase diagram for each compound. We include available experimental data from susceptibility measurements for LiHoF$_4$ (Ref.~\onlinecite{bitko-prl-1996}) and LiErF$_4$ (Ref.~\onlinecite{kraemer-science-2012}) for comparison. Solid and dashed lines of the phase diagram indicate calculations which include and exclude hyperfine interactions, respectively.
\label{fig:liref4:magn}}
\end{figure*}

The knowledge of the CEF parameters provides us with a basis from which we can predict the temperature-field phase diagram of the materials discussed in this paper. To do so we employ a mean-field description wherein many-body interactions are treated as single-ion interactions which sense a self-consistent mean-field. Spin fluctuations are neglected in this treatment and therefore the calculations we shall describe are really the upper limits of critical fields and temperatures.

In order to describe the \LRF\ system we shall employ the Hamiltonian defined in Eq.~\ref{eq:total_Ham_simple}. This can be expressed as,
\begin{eqnarray}
\mathcal{H} &=&\sum_i \left[\mathcal{H}_{\rm CEF}(\mathbf{J}_i) + A \mathbf{J}_i \cdot \mathbf{I}_i- g\mu_B \mathbf{J}_i  \cdot\mathbf{H} \right] \nonumber \\
& & -\frac{1}{2}\sum_{ij} \sum_{\alpha\beta}\mathcal{J}_{\rm D} \overline{\overline{D}}_{\alpha\beta} \mathbf{J}_{i\alpha} \mathbf{J}_{j\beta} - \frac{1}{2}\sum_{\langle ij\rangle}\mathcal{J}_{\rm ex}\mathbf{J}_i \cdot \mathbf{J}_j. \,
\label{eq:ham_detailed}
\end{eqnarray}
The first three single-ion terms describe the CEF, defined in Eq.~\ref{eq:crystal_field}, the hyperfine interaction between electronic and nuclear degrees of freedom with coupling strength $A$ and the Zeeman interaction of the electrons with an applied magnetic field $\mathbf{H}$. The last two interaction terms correspond to dipolar-coupling between moments and super-exchange between nearest-neighbors. As the $4f$ electrons are effectively screened by $5p$ and $5s$ electrons in the outer shell, the exchange interaction is typically very small and has negligible effect on the phase diagram. The long-range nature of dipole-dipole interactions was treated by splitting the dipolar fields' summation into a short-range discrete sum over 100 unit cells and a continuous integration towards the sample boundaries assuming the samples have a spherical shape.\cite{tabei-prb-2008, jensen-book-1991} The dipolar interactions tend to stabilize the system in a magnetically ordered state while a field applied transverse to the axis of the ordered moments acts to cause quantum fluctuations out of the ordered state. The effective Hamiltonian in the $\left| J_z, I_z\right\rangle$ basis can then be diagonalized.

In the preceding section we have used inelastic neutron scattering to derive the CEF parameters for each rare-earth ion. In order to reduce the number of free parameters we have employed the commonly used convention of considering a suitable rotation of the coordinate system in the $ab$ plane such that $B^4_4(s) = 0$. However, now that we are considering a Hamiltonian that includes the dipolar interactions we have to be careful in our choice of coordinate system. The inclusion of these terms means that we no longer have the freedom to rotate our coordinate system. To be rigorous we should map dipolar coupling, Zeeman coupling and so on into the new basis. Unfortunately it is not possible to identify without additional knowledge what the rotation angle about the $c$ axis should be. However, as only $B^4_4(c)$, $B^4_4(s)$ and $B^4_6(c)$, $B^4_6(s)$ parameters, which are found to be small, are affected we have found that the in-plane anisotropy is very small and has only a small effect on the mean-field results we present in this section. We assume that the $x$ axis defined in the coordinate system in \S~\ref{sec:struct_cef} is not far from being along the $a$ axis. Indeed, point charge calculations in LiHoF$_4$ suggest that $x$ axis is only around $-11^\circ$ from $a$ axis.\cite{ronnow-prb-2007}

Figure~\ref{fig:liref4:magn} shows a summary of magnetization data on \LRF. We do not include LiTmF$_4$ as CEF simulations indicate that the system possesses a ground state spin-singlet and we do not expect magnetic order nor a quantum phase transition to develop. We therefore focus on the Gd, Ho, Er and Yb ions which have a doubly degenerate ground state that is split by dipolar-ordered moments at low temperatures (typically below $\approx 2$\,K).

No neutron spectroscopy measurements of LiGdF$_4$ were attempted as the Gd$^{3+}$ ion is not expected to carry any orbital angular momentum. Since the electrostatic interaction between Gd and neighboring ions should not show any dependence on the spin, the crystal environment will not split the degenerate single-ion state. Furthermore, neutron scattering experiments on materials containing Gd are very challenging due to the strongly neutron absorptive nature of Gd isotopes.

Our susceptibility measurements indicate that the system is well described by a Brillouin function between 2 and 300\,K. As expected for a spin-only system, the magnetic susceptibility ($\chi$) of LiGdF$_4$ shows a simple $1/T$-dependence for a measuring field of 5000\,Oe applied along $a$ and $c$ axes, shown in Fig.~\ref{fig:liref4:magn}(a). The high-temperature response is nearly isotropic and is in good agreement with calculations. Fitting the data, we find a Weiss temperature of approximately $+1$\,K which corresponds to antiferromagnetically coupled moments. A small anisotropy is found between $a$ and $c$ components of magnetization of at most 10\%. Although Gd$^{3+}$ is a $S=7/2$ system, other Gd compounds show a far stronger anisotropy than expected for a pure spin system. Such anisotropy can originate from dipole interactions in Gd systems and requires further investigation.\cite{rotter-prb-2003}

As Gd ions are not influenced by the CEF the LiGdF$_4$ system provides an ideal testing ground for pure dipolar interactions at low-temperatures. Mean-field simulations can be used to obtain the magnetic phase diagram of LiGdF$_4$. In the absence of CEF terms, which typically set the energy scales in \LRF, we consider only the dipole-dipole, Zeeman and hyperfine interaction defined in Eq.~\ref{eq:ham_detailed}. Several stable isotopes of Gd exist in naturally occurring Gd with $^{155}$Gd and $^{157}$Gd possessing a nuclear moment $I=3/2$. The hyperfine coupling included in our simulations is very weak -- $^{155}A = 0.031(1)\,\mu$eV and $^{157}A = 0.023(2)\,\mu$eV and would not be expected to affect the phase diagram significantly.\cite{low-pr-1956} We therefore have an ideal system in which to probe the nature of dipolar interactions. In zero field and temperature our calculations quickly converge on bi-layer antiferromagnetic structure with moments in the $ab$ plane, as found in simulations of LiErF$_4$ and LiYbF$_4$. Our simulations predict that LiGdF$_4$ should order below approximately 1\,K (see Fig.~\ref{fig:liref4:magn}(b)). Applying a magnetic field transverse to the spin direction disorders the antiferromagnetic state leading to a quantum phase transition at 0.34\,T, as shown in Fig~\ref{fig:liref4:magn}(c). Figure~\ref{fig:liref4:magn}(d) modeled phase boundary between paramagnetic and antiferromagnetic states in LiGdF$_4$. So far no measurements of the low-temperature properties of LiGdF$_4$ have been reported to verify this result. The hyperfine interaction does not have a significant influence on the phase diagram.

We next discuss the CEF implications on the magnetism in LiHoF$_4$. The low-temperature physics of this system have been extensively studied both experimentally and theoretically.\cite{bitko-prl-1996, chakraborty-prb-2004, ronnow-science-2005, tabei-prb-2008} Ferromagnetism with moments along $c$ is found to develop below 1.53\,K.\cite{bitko-prl-1996} Our measurements of susceptibility from 2 to 300\,K shown in Fig.~\ref{fig:liref4:magn}(e) are in good agreement with calculations based on CEF parameters (Table~\ref{tab:cf_params}). The order parameter as a function of temperature and field are plotted in Figs.~\ref{fig:liref4:magn}(f) and (g), respectively. Based on our model we find $\Tc=1.8$\,K. In the absence of hyperfine coupling we would otherwise expect $\Hc = 3.8$\,T. The inclusion of the hyperfine interaction is found to stabilize magnetic order with a transition into a quantum paramagnetic state above 5.5\,T at zero temperature. In Fig.~\ref{fig:liref4:magn}(h), we show the comparison between the phase diagram obtained from our model and that reported from ac-susceptibility measurements by Bitko~\emph{et al.}~\cite{bitko-prl-1996}. Below approximately $T\approx \Tc/2$ an enhancement of the phase boundary is found which is accounted for by hyperfine coupling of electrons to Ho nuclear moments.

We note that the mean-field calculations for LiHoF$_4$ also includes a renormalization of the $c$ component of the ordered moment due to strong $c$ axis fluctuations such that $\langle J_z\rangle \rightarrow 0.785\langle J_z\rangle$. This gives $\Tc = 1.8$\,K compared to 2.2\,K in the absence of this correction. However, the phase boundary close \Tc\ is still somewhat higher than found experimentally. The result of including the $\langle J_z\rangle$ renormalization is nearly the same as the high-density $1/z$ expansion\cite{ronnow-prb-2007} and is very close to the quantum Monte Carlo simulations (QMC).\cite{chakraborty-prb-2004, tabei-prb-2008} Nevertheless, neither $1/z$ nor QMC theories are able to account for the steep increase of the critical field observed in the phase diagram just below \Tc. Experimentally, the critical exponent is about 0.27 whereas the result of the $1/z$ theory is described by the critical exponent equal to 0.46, i.e. close to 0.5. It remains an open question as to why $1/z$ and QMC models underestimate the role of critical fluctuations as the mean-field approximation is doing.

A detailed study of the nature of magnetic order in LiErF$_4$ has been reported by Kraemer~\emph{et al.}~\cite{kraemer-science-2012}. Neutron powder diffraction was able to refine the magnetic structure to be that of a bi-layered two-dimensional XY antiferromagnet. This is in agreement with our mean-field calculations in the ordered phase. Natural sample of LiErF$_4$ contains 22.8\% of $^{167}$Er which carries a nuclear moment. The hyperfine coupling strength is rather weak 0.5(1)\,$\mu$eV and is found to have an insignificant effect on the phase diagram of LiErF$_4$, shown in Figs.~\ref{fig:liref4:magn}(j)--\ref{fig:liref4:magn}(l). The thermal and quantum phase transitions are found from calculations to be at $\TN = 850$\,mK and $\Hc = 0.95$\,T, respectively. Experimentally, the onset of magnetic order is found at 373(5)\,mK and a critical field of 0.4(1)\,T along $c$ axis is found to drive the system into a quantum paramagnetic state.\cite{kraemer-science-2012} Using CEF parameters obtained from fitting the low-energy CEF excitations ($<25$\,meV) with values reported by Kraemer \emph{et al.}~\cite{kraemer-science-2012}, gives $\TN = 730$\,mK and $\Hc = 0.53$\,T. Therefore, both sets of CEF parameters give critical temperature and fields which are significantly larger than those observed. The mean-field calculation gives a reasonably good description of the qualitative features of the phase diagram. However, LiErF$_4$ shows an unusual character and non-mean-field-like behavior at the phase transitions.\cite{kraemer-science-2012} The dimensional reduction might be related to the frustration of dipolar moments which increase fluctuations. This would imply that the mean-field treatment is unlikely to be very accurate in this system close to the phase boundary.

Up to now, no low temperature magnetization measurements have been reported for LiYbF$_4$. We have performed magnetization measurements along $a$ and $c$ axes between 2 and 300\,K, shown in Fig.~\ref{fig:liref4:magn}(m). We find a relatively large anisotropy between in-plane and out-of-plane susceptibility which is related to the large value of the $B^0_2$ CEF parameter. Our model gives reasonable quantitative agreement except around 100\,K where predicted in-plane susceptibility is somewhat smaller than that measured. We can use our mean-field model and refine CEF parameters to predict the magnetic phase diagram at low temperatures [see Figs.~\ref{fig:liref4:magn}(n)--\ref{fig:liref4:magn}(p)]. We note that there are several isotopes of Yb, some of which carry a nuclear moment. We therefore include in our model a contribution from $^{171}$Yb (14.3\%, $I=1/2$) and $^{173}$Yb (16.1\%, $I=5/2$) of hyperfine coupling strength 11.0 and $-3.0\,\mu$eV, respectively.\cite{flouquet-book-1978} From our calculations we expect to find a transition into a bi-layer antiferromagnetic state with moments along $a$  below 190\,mK. Our simulations predict a magnetic structure which is the same for LiErF$_4$ but with a smaller magnetic moment on the Yb ion. In the absence of hyperfine interaction we would expect $\Hc = 0.51$\,T. However, the interplay of the two hyperfine terms in the Hamiltonian strongly mixes with the electronic degrees of freedom, thereby \Hc\ slowly decreases above 0.7\,T. As shown in Fig.~\ref{fig:liref4:magn}(p), the effect is most pronounced below 20\,mK, while above this temperature hyperfine interaction has the converse effect of destabilizing magnetic order and reducing the field at which the sample becomes a quantum paramagnet.

\section{Conclusion}

We have performed inelastic neutron scattering measurements of \LRF\ where $R =$ Ho, Er, Yb and Tm systems to analyze the low-energy CEF excitations. The understanding of the crystal level scheme is important in providing a handle on the magnetic properties of \LRF\ which exhibit intriguing behavior at low temperatures due to dipolar interactions. We find that our data is in good agreement with previous spectroscopic studies. Using the extracted CEF parameters we make qualitative predictions using mean-field calculations of the magnetic phase diagrams of \LRF\ which can be tested against measurements and aid in the pursuit of understanding fundamental dipolar interactions. We present minimal models based on electron-phonon coupling to explain CEF linewidth broadening in LiTmF$_4$ and splitting of the first excited CEF state in LiYbF$_4$. Such coupling can have an effect on many different physical quantities such as lattice parameters, thermal expansion, magnetostriction as well as phonon spectrum. Using high-resolution neutron scattering measurements on LiHo$_x$Y$_{1-x}$F$_4$, $x = 0.045$ we show from RPA calculations that the hyperfine splitting of the doublet ground state can account for the CEF line asymmetry and width. However, in LiHoF$_4$, we find anomalous broadening of the same CEF level which could be an indication of magnetoelastic coupling.

\begin{acknowledgments}
We are grateful to A.~T. Boothroyd for many insightful discussions. We wish to thank P. Huang for his help in setting up the SQUID measurements. M.J. is grateful for support from Marie Curie Action COFUND (EPFL Fellows). This work was supported by ERC project CONQUEST and the Swiss National Science Foundation. Neutron scattering experiments were performed at Paul Scherrer Institute, SINQ (Switzerland), Forschungs-Neutronenquelle Heinz Maier-Leibnitz, FRMII (Germany), Rutherford Appleton Laboratory, ISIS (UK) and  Argonne National Laboratory, IPNS (USA).
\end{acknowledgments}

\bibliographystyle{apsrev4-1}

\bibliography{shorttitles,biblio}

\end{document}